\documentstyle[12pt]{article}

\def\psibar{{\bar \Psi}}

\begin{document}

\begin{titlepage}
\begin{flushright}
\hfil                 UCSD/PTH 92-29\\
\hfil August 1992 \\
\end{flushright}
\begin{center}

\vspace{1cm}
{\LARGE\bf Critical Momenta of Lattice Chiral Fermions }

\vspace{2.0cm}
{\large Karl Jansen and Martin Schmaltz}\\
\vspace{1cm}
University of California at San Diego\\
Department of Physics-0319 \\
9500 Gilman Drive\\
La Jolla, CA~92093-0319 \\
USA  \\
\vspace{0.2cm}

\vspace{2.0cm}
\end{center}

\abstract{
\noindent We determine the critical momenta for chiral fermions in
the domain wall model recently suggested by Kaplan.
For a wide range of domain wall masses $m$ and Wilson couplings $r$
we explicitly
exhibit the regions in momentum space where the fermions are chiral.
We compare the critical momenta for the
infinitely large system with those obtained on a finite lattice.
}
\vfill
\end{titlepage}

\section{Introduction}

In a recent paper \cite{david} a new method for simulating chiral
fermions on a lattice was suggested.
The proposal uses the fact that by introducing a
domain wall in an odd dimensional --~and therefore vectorlike-- theory one
finds zero modes bound to this domain wall \cite{jarebbi,gw,caha}.
Regarding the domain
wall as a lower (even)
dimensional world it was demonstrated in \cite{david}
that for the infinite lattice
these zero modes are chiral fermions
and that the resulting theory on
the domain wall exhibits the desired anomaly structure.

The same features
have been shown to survive on a finite lattice where
boundary conditions require one to consider a wall/anti-wall pair.
For low energies a chiral fermion is
bound to one of the domain walls with its chiral partner of
opposite chirality living on the other domain wall.
In addition, it has been demonstrated
that in the presence of weak external gauge fields the divergence of the
gauge current satisfies the {\em continuum} anomaly equation
\cite{karl}.
This surprising result has been explained as being a consequence of the fact
that on the lattice the
divergence of the Goldstone-Wilczek current \cite{gw,caha}
has the same form
as in the continuum \cite{gojaka}.

One of the most important properties of the domain wall method is
that the fermions are only chiral in the low energy limit.
For the special choice of the Wilson coupling $r=1$ it was shown
in \cite{david} that there exists a critical value of the momentum
$p_c$ for which the fermions cease to be chiral.
However, it is expected that also for other choices of the Wilson coupling
such critical momenta exist.
This was confirmed on the finite lattice where
the existence of the critical momentum could be demonstrated for an
$r=1.8$ \cite{karl}.

Here we want to perform a systematic study of the values of the critical
momenta for which one will find chiral modes. We compare the analytical
results obtained on the infinite lattice with
the spectrum of the finite lattice Hamiltonian.
Aside from the theoretical interest of this question, our results are of
practical importance for future numerical simulations of this system.

\section{Zeromodes on the Infinite Lattice}

To be specific we will first discuss the zeromodes for the case of a
3-dimensional model, though our results will be generalized to
arbitrary dimensions at the end. We start with the Dirac-Wilson operator
on an infinite lattice with lattice spacing $a=1$

\begin{equation}
K_{3D}  =\sum_{\mu=1}^3\sigma_\mu\partial_\mu
+m\theta(s)+\frac{r}{2}\sum_{\mu=1}^3\Delta_\mu
\label{eq:diracco}
\end{equation}
\noindent where $\partial$ denotes the lattice derivative
$\partial_\mu = \frac{1}{2}\left[\delta_{z,z+\mu} - \delta_{z,z-\mu}\right] $,
$\Delta$ the lattice Laplacian
$\Delta_\mu = \left[\delta_{z,z+\mu} + \delta_{z,z-\mu}-2\delta_{z,z}\right] $,
the $\sigma_\mu$ are the usual Pauli matrices and
$r$ the Wilson coupling.
We will denote by $s$ the extra dimension along which the mass defect
appears, while $x,t$ are the 2-dimensional coordinates.
The domain wall is taken to be a step function $\theta$,
\begin{equation}
\theta(s) = \left\{ \begin{array}{lll}
                    -1 & s < 0 \\
                     0 & s=0\\
                    +1 & s > 0
                     \end{array} \right.
\label{eq:theta}
\end{equation}
\noindent where the height of the domain wall is given by the mass parameter
$m$ which we will choose to be positive throughout this paper.

We are looking for solutions which are plane waves in the $(x,t)$-plane
\begin{equation}
\Psi^{\pm} = e^{i(p_tt+p_xx)}\Phi(s)u^{\pm}
\label{eq:psi}
\end{equation}
where $u^\pm$ are the eigenspinors of $\sigma_3$
\begin{equation}
\sigma_3 u^\pm = \pm u^\pm  .
\label{eq:spinors}
\end{equation}
With this ansatz the Dirac operator becomes
\begin{equation}
K_{3D}  =\sum_{i=1}^2i\sigma_i\sin(p_i)+ \sigma_3\partial_s
+m\theta(s)+r(\sum_{i=1}^2(\cos(p_i)-1)+\frac{r}{2}\Delta_s  \;\; .
\label{eq:diracmom}
\end{equation}
Our final goal is to diagonalize the 3 dimensional Dirac operator
in such a way that it reduces to the 2 dimensional Dirac operator
for free massless fermions,
\begin{equation}
K_{3D}\Psi = K_{2D}\Psi
\end{equation}
where $K_{2D}$ acting on $\Psi$ is given by
\begin{equation}
K_{2D} = \sum_{\mu=1}^2\sigma_\mu\partial_\mu = i(\sigma_1 \sin(p_t)
+\sigma_2 \sin(p_x))\;.
\label{eq:e1}
\end{equation}
\noindent Hence the equation to solve is
\begin{equation}
\left[\sigma_3\partial_s + m\theta(s) -rF
+\frac{r}{2}\Delta_s\right]\Phi u^\pm =0
\label{eq:problem}
\end{equation}
where
\begin{equation}
F=\sum_{i=t,x}(1-\cos(p_i)) .
\label{eq:F}
\end{equation}
Following \cite{david} we choose an exponential ansatz for $\Phi$
away from the domain wall
\begin{equation}
\Phi(s+1)=z\Phi(s) .
\label{eq:ansatz}
\end{equation}
Inserting this into (\ref{eq:problem}) one finds four solutions
\begin{equation}
z =\frac{r-m_{eff}\pm\sqrt{m_{eff}(m_{eff}-2r)+1}}{r\pm1}
\label{eq:roots}
\end{equation}
where $m_{eff} = m\theta(s)-rF$, the $\pm$
in the nominator stand for the two roots and the $\pm$ in the
denominator stand for the chirality.
Note that in the limit $r=1$ eq.(\ref{eq:roots}) can be reduced to the
corresponding expressions in \cite{david}.

We have to impose the condition that the solutions are normalizable to
obtain sensible wavefunctions. This means
that
$|z|>1$ for $s<0$ and $|z|<1$ for $s>0$.
The boundaries of the regions where chiral solutions exist are obtained
by setting $|z|=1$.
Explicit matching of the normalizable solutions for
positive and negative $s$ at $s=0$
enables us to determine the regions with chiral fermions.
We find that existence and chirality of the solutions is independent of
the sign of $r$
and that a negative $m$ leads to opposite chiralities.
Depending on the values of $m/r$ we get
$m =rF$ and $m =r(F+2)$ as the boundaries for the critical momenta
where $F$ is defined as above in eq.(\ref{eq:F}).

The results are summarized
in fig.1. We show the Brillouin zone $-\pi\le p_t \le \pi$, \hbox{$-\pi \le p_x
\le \pi$} for different ratios of $m/r$. The white areas indicate the
region in momentum space where chiral fermions exist.
Starting with $m/r=0$ we find no chiral fermions. For increasing $m/r>0$
the region where chiral modes exist grows from a small circle around
$\vec{p} =(0,0)$ until it hits the boundary of the
Brillouin zone for $m/r =2$. The boundary of the circle, i.e. the upper
critical momenta, is given by $m=rF$.
Increasing $m/r$ further opens up the two ``doubler''
modes at $\vec{p} =(0,\pi)$ and $\vec{p} =(\pi,0)$ which have flipped
chirality, while the original mode at $\vec{p} =(0,0)$ disappears.
Here the boundaries of the white regions are given by $m=rF$ for the
lower and $m=r(F+2)$ for the upper critical momenta.
For $m/r>4$ the two ``doublers'' disappear and we get a zero mode at
$\vec{p} =(\pi,\pi)$ with the same chirality as the mode
at $\vec{p} =(0,0)$.
The boundary for the lower critical momenta is
given by $m=r(F+2)$. This mode is finally
also lost as $m/r$ is increased to $m/r \ge 6$.

%
%
%

We want to remark that this spectrum stems from $\Psi^+$
solutions only, and that
there are no $\Psi^-$ solutions for positive $m$. The
change of the chirality is the usual reinterpretation of the
chirality at different corners of the Brillouin zone.

The
generalization to arbitrary dimensions $d=2n+1$
consists merely in replacing the
function $F$ in eq.(\ref{eq:F}) by
\begin{equation}
F=\sum_i^{d-1} (1-\cos(p_i))\;\;.
\label{generalf}
\end{equation}
We find in $d$ dimensions
that for $2k <|m/r|<2k+2$ ($0\le k <d-1$), the
number of chiral zero modes $N_{zm}$ bound to the $d-1$ dimensional
domain wall is given by
\begin{equation}
N_{zm}= \left( \begin{array}{c}
 d-1 \\ k \end{array} \right)
\label{eq:nzm}
\end{equation}
and their chirality is $(-1)^k$.

The regions of chiral fermions as found here correspond exactly to the
values of $m/r$ where the lattice Chern-Simons current
induced by Wilson fermions
changes its value \cite{gojaka} giving contributions to the Goldstone-Wilczek
current.
The calculation of the Chern-Simons current in \cite{gojaka}
uses the observation that the fermion
propagator in momentum space can be interpreted as
a map from the torus to the sphere ($T^{d}\rightarrow S^{d}$).
The winding number of this map is closely related to the zeromode spectrum
and changes only at particular values of $m/r$ which agree exactly
with our results for the values of $m/r$ where the number of
zeromodes changes.

\section{Finite Size Effects}

Any numerical work on this system will necessarily involve finite lattices,
and so we now compare the critical momenta obtained on the infinite
system with the ones of a finite lattice.
On the finite lattice we have to choose
some boundary conditions which generate a second anti-domain wall.
The mass term is therefore modified to be $m\theta_L(s)$ with
\begin{equation}
\theta_L(s) = \left\{ \begin{array}{lll}
                    -1 & 2\le s \le \frac{L_s}{2} \\
                    +1 & \frac{L_s}{2}+2 \le s \le L_s \\
                     0 & s=1, \frac{L_s}{2}+1
                     \end{array} \right. \;\; .
\label{eq:m0}
\end{equation}

We searched for the zeromodes on the finite lattice by solving the
Hamiltonian problem numerically.
If we again assume plane waves in the $x$-direction the
Hamiltonian is given by \cite{karl}
\begin{equation}
H  =-\sigma_1\left[i\sigma_2\sin(p_x) + \sigma_3\partial_s
+m\theta_L(s)+r(\cos(p_x)-1) +\frac{r}{2}\Delta_s) \right] \;\; .
\label{eq:hamiltonian}
\end{equation}

We computed the eigenvalues and eigenfunctions of the Hamiltonian
eq.(\ref{eq:hamiltonian}) numerically~\footnote
{In principle it is, of course,
possible to perform a similar analysis as in the infinite system.
However, on the finite lattice one also has to take into account
solutions which are non-normalizable in the
infinite system. In addition, on the
finite lattice the modes are not exactly chiral. A mode bound to one of
the walls always gets an exponentially suppressed  contribution from the mode
on the other wall. These features render analytic calculations on the finite
lattice quite intractable.}.
Choosing the Hamiltonian instead
of the Dirac
operator reduces the numerical effort substantially as we only have to search
for a single critical momentum instead of a pair.
To find the critical momenta we studied the ratio
\begin{equation}
R=\frac{\psibar\Psi}{\psibar\sigma_1\Psi}
\label{eq:R}
\end{equation}
which is a normalized measure for whether the fermions are chiral or not. It is
zero
if the fermions are chiral and $R>0$ for non-chiral modes (see
fig.2b in \cite{karl}). To determine whether we still have chiral
fermions we defined a threshold value for R. If $R < 0.01$ we regarded
the fermions to be chiral.

We show the critical momenta as functions of $r$ at three different
values of $m$ in Fig.2. As discussed above we should find two boundary
curves for the critical momenta.
Note that because we now use the Hamiltonian, the function $F$ is
only one dimensional $F=1-\cos(p_x)$.
For $m/r<2$ the chiral fermions appear
for momenta bounded by $m=rF$. This is the solid line in
Fig.2. For $2<m/r<4$ the momenta for which chiral fermions appear are
given by $m=rF$ for the lower and $m=r(F+2)$ for the upper critical
momenta. We plot the curve for the upper critical momenta as a
dashed line in Fig.2.

We compare the results from the infinite system
with our finite lattice calculations. The crosses correspond to a system
size of $L=100$ and the open circles to a size of $L=20$. We find that the
$L=100$ lattice is practically indistinguishable from the infinite
system. For $L=20$, a lattice size realistic for simulations, a
small shift occurs. Fixing $m$ and $r$ we find for $m/r<2$ a smaller value
and for $2<m/r<4$ a larger value of the critical momentum. Note,
that for $m=0.4$ the circles belong only to the solid curve. We did not
find the momenta which correspond to the dashed line as our resolution
in the numerical computation was not fine enough.

In conclusion, we find the differences
between $L=\infty$ and $L=20$ to be small.
This means that the lattice has not to be too large to
reproduce the basic features of
the model at $L=\infty$ which
makes the domain wall model feasible for numerical
investigations.
Therefore we have shown that the domain wall model can
be used for numerical simulations on realistic lattices. We also give
the values of the domain wall mass $m$ and the Wilson
coupling $r$ with which numerical simulations should eventually be performed.

%
%
%

\section*{Acknowledgements}
We want to thank J. Kuti and D. Kaplan for numerous helpful
discussions.
This work is supported by DOE grant DE-FG-03-90ER40546.
\pagebreak

\pagebreak

\section*{Figure Caption}

{\bf Fig.1} We plot the regions within
the Brillouin zone \hbox{$-\pi<p_t<+\pi$}, \hbox{$-\pi<p_x<+\pi$}
where chiral fermions exist (white areas)
as a function of $m/r$ with $m$ the domain wall mass and $r$ the Wilson
coupling. For the different cases $0<m/r<2$,
$2<m/r <4$ and $4<m/r<6$ we have one chiral fermion with positive
chirality, two chiral fermions with negative chirality and again one
chiral fermion with postive chirality, respectively. For $m/r<0$ and
$m/r>6$ there exist no chiral fermions.

\noindent {\bf Fig.2}
We plot the two lines which give the upper and lower
critical momenta on the infinite system,
$m=rF$ (solid lines) and  $m=r(F+2)$ dashed lines, where
$F=1-\cos(p_x)$. We show these lines at three different values of the
domain wall mass $m$ as a function of the Wilson coupling $r$. We compare
the curves from the infinite system with results from finite lattice
calculations with lattice sizes $L=100$ (crosses) and $L=20$ (open
circles). For $m=0.4$ our resolution in the numerical
computations was not fine enough to find the critical momenta
corresponding to the dashed line. Note that though a shift in the
critical momenta on the finite lattice is visible we find the same
structure as for the infinite system.

\end{document}